\documentclass[
reprint,
 amsmath,amssymb,
 aps,
]{revtex4-2}

\usepackage{hyperref}% add hypertext capabilities

\usepackage{graphicx}
\usepackage{siunitx}

\begin{document}

\title{Highly-linear flux-to-voltage transducer based on superconducting quantum interference proximity transistors}
% Force line breaks with \\
\author{Angelo Greco}
\affiliation{NEST, Istituto Nanoscienze-CNR and Scuola Normale Superiore, Piazza S. Silvestro 12, I-56127 Pisa, Italy}
 
\author{Giorgio De Simoni}
\email{giorgio.desimoni@nano.cnr.it}
\affiliation{NEST, Istituto Nanoscienze-CNR and Scuola Normale Superiore, Piazza S. Silvestro 12, I-56127 Pisa, Italy}

\author{Francesco Giazotto}
\affiliation{NEST, Istituto Nanoscienze-CNR and Scuola Normale Superiore, Piazza S. Silvestro 12, I-56127 Pisa, Italy}

%\date{\today}

\begin{abstract}
Superconducting quantum interference devices (SQUIDs) are state-of-the-art in ultra-sensitive magnetometry; however, conventional SQUID devices are fundamentally limited by the inherently nonlinear and periodic nature of their transfer function. Although flux-locked loop (FLL) configurations can mitigate this issue, they introduce electronic complexity and bandwidth constraints that hinder scalability in quantum circuits. In this work, we present an experimental demonstration of the bi-SQUIPT, a flux transducer that modulates the density of states in a proximitized superconducting weak link. The device employs a dual-loop architecture with differential readout, which enables cancellation of non-linearities typical of individual elements, achieving a voltage swing of approximately 120 $\mu$V. Measurements yield a spurious-free dynamic range (SFDR) of up to 60 dB, consistent with theoretical predictions and comparable to that of SQUID arrays, while maintaining power dissipation in the femtowatt range. The results further highlight a remarkable operational stability up to 600 mK, positioning the bi-SQUIPT as an enabling technology for high-density cryogenic quantum electronics.
\end{abstract}


\maketitle

\section*{Introduction}
Superconducting interferometers are state-of-the-art for ultra-sensitive magnetometry and cryogenic signal processing. Since their inception, superconducting quantum interference devices (SQUIDs) \cite{Clarke2004} have approached the quantum limits of sensitivity, enabling applications ranging from magnetic bio-imaging \cite{VRBA2001249, Makela2016_SQUIDsInBiomagnetism, Oyama2006_DigitalFLLforSQUID, Zevenhoven2020_SQUID_receiver_arrays} to the readout of cryogenic quantum devices. However, standard SQUIDs suffer from a fundamental limitation: their flux-to-voltage transfer function $V(\Phi)$ is inherently periodic and quasi-sinusoidal. This non-linearity severely restricts their dynamic range and introduces harmonic distortion when handling relatively large signals. To mitigate this, SQUIDs are conventionally operated in a flux-locked loop (FLL) configuration \cite{Drung2003_HighTcLowTc_dcSQUID_electronics, Drung2006_SQUIDReadoutElectronics, Oyama2006_DigitalFLLforSQUID, 4277368}, which keeps the SQUID locked at a fixed operating point by applying a feedback flux that cancels any incoming flux, so the measured signal becomes the feedback current needed to maintain this balance. Although effective for low-frequency applications, FLL electronics impose bandwidth limitations, introduce additional noise, and entail significant complexity, thus hindering the scalability required for large-scale quantum circuits \cite{4277368}. Consequently, intrinsically linear superconducting flux-to-voltage transducers with low power dissipation must be regarded as an enabling technology for a multitude of applications, and their development should be considered a key research frontier.

Several architectures have been proposed to engineer linearity directly into the physical response of the device. A prominent approach is the bi-SQUID, initially proposed by Kornev \textit{et al.} \cite{kornevBiSQUIDNovelLinearization2009, kornevBiSQUIDDesignApplications2020, kornevHighInductanceBiSQUID2017, kornevHighLinearityBiSQUIDDesign2018, kornevSignalNoiseCharacteristics2014}. The bi-SQUID modifies the standard direct-current (DC) SQUID topology by inserting a third Josephson junction that acts as a non-linear inductive shunt. This additional element reshapes the effective current-flux characteristics of the interferometer, transforming the sinusoidal response into a quasi-triangular waveform with linearity capable, in principle, of exceeding 120 dB in total harmonic distortion (THD) \cite{kornevBiSQUIDNovelLinearization2009}. However, single-interferometer experimental realizations have struggled so far to match these theoretical predictions, saturating below $\sim50$ dB \cite{PhysRevApplied.18.014073, Trupiano2025_HighlyLinear_BiSQUID} both in spur-free dynamic range (SFDR) and THD due to stringent requirements on fabrication symmetry. 
Recent advances have attempted to address these fabrication variances: an electrostatic gate-controlled Josephson weak-link \cite{WO2023081970, Kong2024_BiSQUID_gate_control, DeSimoni2018_MetallicSupercurrentFieldEffectTransistor, 10.1063/5.0136709, DeSimoni2021_GateControl_CurrentFluxRelation, Paolucci2019_FieldEffectControllableMetallicJosephsonInterferometer} was proposed as a tool to control the critical current of the shunting junction of a bi-SQUID via gate voltage. Although such tunability offers a promising pathway to compensate for parameter mismatch and to recover high linearity post-fabrication, bi-SQUIDs typically operate in the nano to micro-watt dissipation regime, thereby imposing a significant thermal load on dilution refrigerators operating at temperatures below 100 mK.

In parallel with geometrical modifications, differential configurations, known as DSQUIDs (Double SQUIDs), have been explored to improve linearity and sensitivity \cite{Soloviev2019_DifferentialDCSQUID}. The D-SQUID uses two SQUIDs in a differential readout scheme that, in addition to suppressing common-mode noise, enables subtraction of part of the SQUID voltage response from its mirrored image, thereby partially compensating for nonlinearities in the magnetic flux-to-voltage transfer function.

This concepts finds their ultimate expression in superconducting quantum arrays (SQAs)\cite{Kornev2017_BroadbandSuperconductorAntennas, Kornev2014_SuperconductingQuantumArrays_TAS, Kornev2014_SuperconductingQuantumArrays_Conf, Kornev2013_ActiveSQArrayAntenna, Kornev2014_ParallelArrayCell, Kornev2011_SQIFArrays_Design, Kornev2010_HighLinearityJosephson, Kornev2009_HighLinearitySQIF, Kornev2009_PerformanceSQIFs}, which offer high linearity at the cost of large physical footprints and high power consumption scaling with the number of elements of the array, making them unsuitable for dense integration at the millikelvin stage.

In this context, the superconducting quantum interference proximity transistor (SQUIPT) offers a paradigm shift \cite{Giazotto2010, GiazottoTaddei2011,PhysRevB.84.214514, Ronzani2014,10.1063/1.4866584, DAmbrosio2015, Virtanen2018, Jabdaraghi2017, Virtanen2016, Jabdaraghi2018,Paolucci2022, Strambini2016, Ligato2022, Ligato2021}. Unlike SQUIDs that rely on the phase-dependent supercurrent in Josephson junctions, the SQUIPT exploits the flux-controlled modulation of the quasiparticle density of states (DOS) in a superconducting weak-link, consisting of a normal-metal (N) or superconducting (S) constriction \cite{Virtanen2016, Ligato2022, Ligato2021}, closed in a superconducting ring. Indeed, the Josephson effect \cite{Likharev1979} occurs in an S-type \cite{Vijay2010, LevensonFalk2013}  or  N-type weak link \cite{GiazottoTaddei2011, Giazotto2004, Carillo2006, Savin2004, DeSimoni2021_GateControl_CurrentFluxRelation,DeSimoni2019} in clean galvanic contact with S leads as a consequence of the so-called superconducting proximity effect. The latter stems from the formation of Andreev bound states in the weak link \cite{Pannetier2000, Belzig1999, McMillan1968}, which open a minigap in the DOS, with an amplitude that depends on the phase drop $\delta$ across the weak link. The magnetic flux $\Phi$ threading the S ring controls $\delta$ through the flux quantization relation, and it can be quantified by measuring the current (I) versus voltage (V) characteristics of a superconducting or normal-metal tunnel electrode tunnel-coupled to the weak link by a thin insulating barrier. Indeed, the tunnel current is closely determined by the phase-dependent DOS in the weak link.
SQUIPTs built with different architectures \cite{Giazotto2010,DAmbrosio2015,Strambini2016,Ronzani2017, Meschke2011} demonstrated a flux sensitivity slightly above the theoretical flux-noise boundary of $n\Phi_0 / \sqrt{Hz}$ \cite{GiazottoTaddei2011,Meschke2011,Ronzani2017} and reduced power dissipation due to the high impedance ($10^4 \div 10^6$ $\Omega$) of the tunnel probe.
However, SQUIPTs exhibit a highly nonlinear voltage-to-flux response; therefore, to bridge the gap between the low dissipation of SQUIPTs and the linearity of bi-SQUIDs and DSQUIDs, bi-SQUIPTs \cite{desimoni2023} were theoretically proposed. 
These devices employ a differential-readout dual-loop architecture, similar to a DSQUID, in which the DOS modulation of two SQUIPTs is combined to cancel non-linearities, yielding a predicted SFDR of $\sim 60$ dB while maintaining femtowatt-level power dissipation.

Here, we report the first experimental demonstration of the bi-SQUIPT. We provide a comprehensive characterization of its magneto-electric response, verify the predicted linearization mechanism, and demonstrate its superiority as a low-dissipation, high-linearity flux-to-voltage transducer for cryogenic quantum electronics.

\section{Single SQUIPT characterization}
In the initial proposal \cite{desimoni2023}, the bi-SQUIPT was defined as a three-terminal double-loop superconducting interferometer, consisting of two SQUIPTs sharing one of the two loop arms. Each SQUIPT includes a superconducting ring closed on a Josephson weak-link, tunnel-coupled with a superconducting or normal-metal tunnel electrode by a thin insulating barrier through which a current can flow, whose amplitude is determined by the phase-dependent DOS in the weak-link. Since flux quantization holds independently in the two loops even though they share one arm, a bi-SQUIPTs is, from a circuital point of view, fully equivalent to two independent SQUIPTs (from now on, labeled a and b, respectively) connected in parallel and sharing a common ground on the loops with the voltage drops across the tunnel junctions ($V_a$ and $V_b$) being measured differentially ($V=V_{a}-V_{b}$). It should be noted that, since the bi-SQUIPT is meant to sense a magnetic flux common to both the loops, for perfectly identical SQUIPTs, $V$ is by construction always equal to zero, unless an additional flux is fed through an \textit{ad hoc} flux line, to one of the loops. Moreover, although from an application point of view the original proposal allows greater compactness and is more suited for applications requiring high spatial resolution of the external flux (like, \textit{e. g.}, scanning-SQUID magnetometry), for this work we decided to adopt the second configuration, because it allows separate characterization of each tunnel junction and consequently to select those that are more similar. In fact, the maximum linearity of the voltage response is expected for two SQUIPTs with identical areas, tunnel resistances, and bias currents; even if imbalances in the tunnel resistances of the two tunnel probes can be corrected by driving the device with two suitably different bias currents, which balance the amplitudes of the voltage outputs of the individual SQUIPTs and, consequently, of the bi-SQUPIT as a whole. In the tunnel limit, the latter can be obtained by assuming a negligible Josephson current due to a suitably high tunnel resistance, inverting the usual quasiparticle tunneling current relation independently for the two SQUIPTs \cite{GiazottoTaddei2011,desimoni2023}:
\begin{equation}
\label{eq:IV}
\begin{split}
I_{b,x}(V_{x})=\frac{1}{ewR_{x}}\times
\\
\int_{\frac{L-w}{2}}^{\frac{L+w}{2}}dl\int_{-\infty}^{\infty}d\varepsilon \mathcal{D}_{w,x}(l,\varepsilon,\varphi_{x},T) \times
\\
\mathcal{D}_{p,x}(\tilde{\varepsilon}_{x},T)F(\varepsilon,\tilde{\varepsilon}_{x},T),
\end{split}
\end{equation} 
where $\varepsilon$ is the energy relative to the chemical potential in the superconductor, $\tilde{\varepsilon}_{x}=\varepsilon-eV^{x}$, $F(\varepsilon,\tilde{\varepsilon}_{x})=[f_0(\tilde{\varepsilon}_{x})-f_0(\varepsilon)]$, $f_0(\varepsilon)$ is the Fermi-Dirac energy distribution function, $R_{x}$ is the normal-state tunneling resistance of the $x$ probing junctions, $l \in[0, L]$ is the spatial coordinate along the weak-links,  $w$ is the width of the probing junctions, and $L$ is the length of the proximitized weak links. $\mathcal{D}_{w,x}$ and $\mathcal{D}_{p,x}(\varepsilon,T)=|\Re[(\varepsilon+i\Gamma)/\sqrt{(\varepsilon+i\Gamma)^2-\Delta^2(T)}]|$ are the DOSs at electronic temperature $T$ of the S probes and of the weak-link, respectively, where $\Delta(T)$ accounts for the temperature evolution of the superconducting order parameter with respect to its zero-temperature value $\Delta_0$, $T_C$ is the critical temperature of the superconductor, and $\Gamma$ is the Dynes parameter that describes the phenomenological DOS broadening. $\mathcal{D}_{w,x}$ contains the flux-driven phase-dependent behavior of the induced minigap in the weak-link and can be analytically evaluated in the short-junction limit only, as fully discussed in \cite{desimoni2023}. Although our devices, which are Al superconducting loops closed on a Cu weak-link tunnel-coupled to an Al/Al-oxide probe, do not strictly fall into the short-junction case, they are still satisfactorily described by this model.

\begin{figure*}[t]
    \centering
    \includegraphics[width=0.8\textwidth]{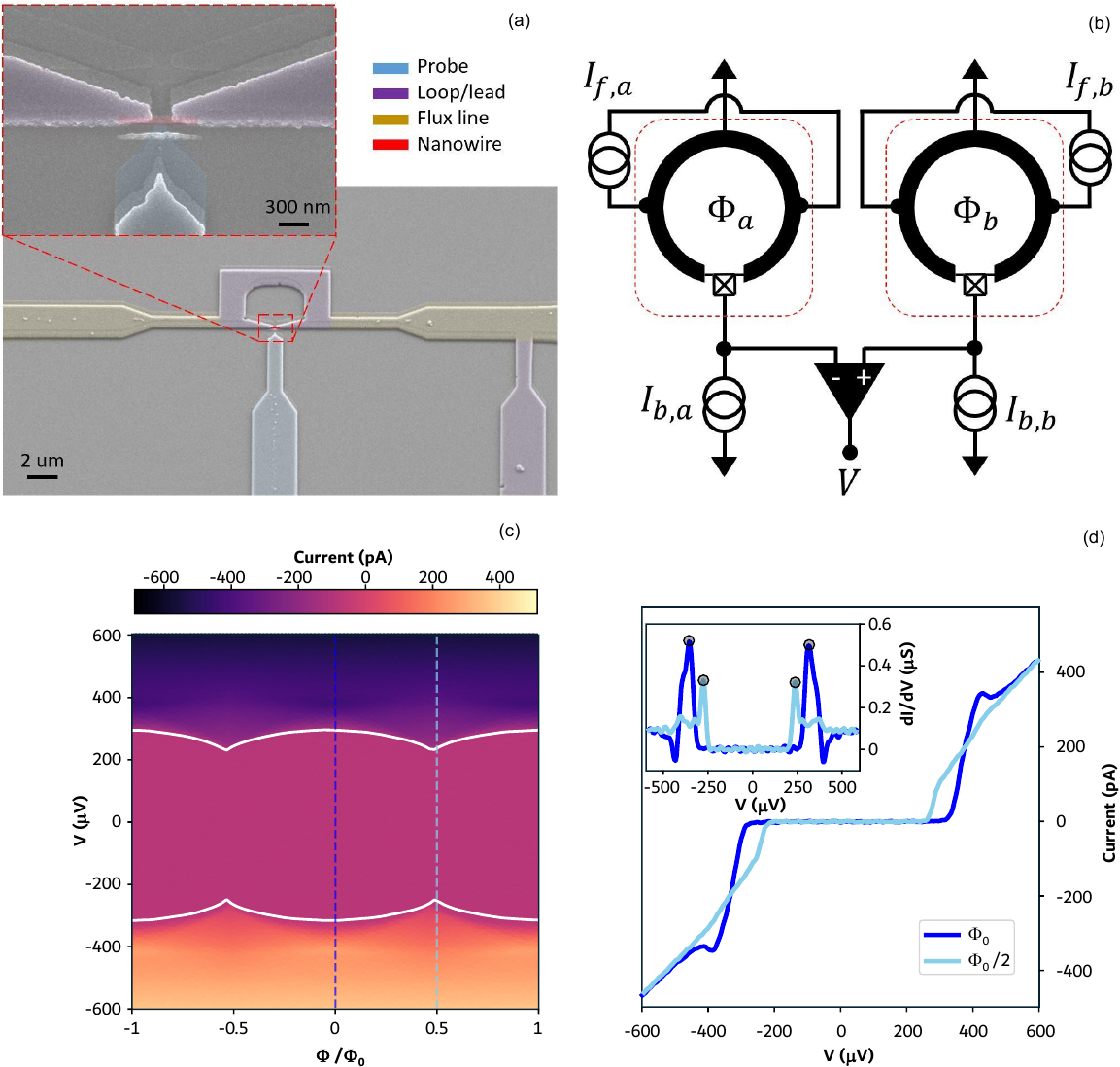}
    \caption{\textbf{Circuit and IV characterization of single SQUIPT:} \textbf{(a)} False color micrograph of a SQUIPT. The device consists of a tunnel-probe lead (blue), flux lines (yellow), a loop and a ground lead (purple), and a nanowire (red). \textbf{(b)} Circuit schematic of a bi-SQUIPT. Two SQUIPTs are connected in parallel, thereby grounding the loop and enabling measurement of the voltage drop between their tunnel probes. Each SQUIPT has a current bias that is used to set it at the correct point on the IV curve to maximize the voltage swing. 
    Different magnetic fluxes pierce both loops. \textbf{(c)} IV curves of a typical SQUIPT as a function of the magnetic flux in its loop. \textbf{(d)} Cuts of the color plot shown in (c) at $\Phi = \Phi_0$ and $\Phi = \Phi_0/2$.}
    \label{fig:Fig1}
\end{figure*}
The SQUIPTs were fabricated using shadow mask evaporation in an electron beam evaporator with a base pressure of $3\cdot10^{-11}\:$Torr at three different angles. In the first evaporation, to realize the tunnel probe, we deposited $20\:$nm of Al at an angle of 25°, after which we proceeded with a thermal oxidation of 10 minutes at $7.3\:$Torr. The second and third depositions were performed in rapid succession to ensure clean contact between the normal-metal Cu nanowire and the Al superconducting leads. The $20\:$nm-thick Cu layer was deposited at a 0° angle, while the $200\:$nm-thick Al leads were deposited at 16°. In \autoref{fig:Fig1}(a), a false color micrograph of a representative SQUIPT is shown. In blue and purple, we highlight the tunnel probe and loop lead, respectively. In yellow, we show a galvanically connected flux line to the loop having an area of $\sim$ \SI{4}{\micro\meter^2}. The flux line induces magnetic flux in each loop via its self-inductance. The inset of \autoref{fig:Fig1}(a) shows a higher-magnification scanning-electron micrograph of the same device of \autoref{fig:Fig1}(a), in which the \SI{210}{\nano\meter} long Cu weak-link is false-colored in red. The intersection of the probe and the weak link defines the tunnel-junction area. Panel (b) shows a schematic of the wiring setup of the bi-SQUIPT in the current bias configuration. The red dashed squares enclose the two SQUIPTs that, operated in parallel, form the bi-SQUIPT. Each SQUIPT is biased by a constant current flowing through the tunnel probe $I_{b,x}$ (with $x=a,b$). A current $I_{f, x}$ is fed into the loop through the flux lines, resulting in an induced magnetic flux $\Phi_x$ in the loop area. Both loop leads are connected to a common ground and the output voltage $V$ is measured differentially between the two tunnel probes. 
Our setup also includes a superconducting electromagnet that is used to thread a uniform static magnetic field common to both devices.

To separately characterize the performance of each SQUIPT forming the bi-SQUIPT, we measured their IV curves as a function of the magnetic flux piercing the loops. This was achieved by voltage-biasing the SQUIPT and measuring the device current as a function of the magnetic field induced by the superconducting coil. A typical result is shown in \autoref{fig:Fig1}(c), where $I_{b,x}$ is color plotted as a function of the voltage bias and the magnetic flux $\Phi$. The area in magenta, enclosed between the white lines, represents the subgap region, hence where the bias voltage is lower than $\left(\Delta_{probe}+\Delta_{wire}(\Phi)\right)/e$, with $\Delta_{probe}$ and $\Delta_{wire}(\Phi)$ respectively the superconducting gaps of the Al tunnel probe and the proximitized Cu wire, and $e$ is the elementary charge. $\Delta_{wire}(\Phi)$ is a function of the magnetic flux piercing the interferometer; indeed, a modulation of the magenta area with $\Phi$ is apparent in the plot, corresponding to the modulation of the subgap edge. In panel (d) of  \autoref{fig:Fig1}, we show two line-cuts extracted from panel (c), taken at $\Phi_0$ and $\Phi_0/2$ (corresponding to dark and light blue dashed lines, respectively). These IVs show modulation of the subgap region, which is at its extremal points for these specific flux values. In the inset, we also plot the differential conductance $\frac{dI}{dV}$, calculated through the numerical derivative of the corresponding IV curves, which allow us to estimate the gap values by evaluating the energy difference ($e \Delta V$) between the apical point of the differential conductance [black dots in the inset of \autoref{fig:Fig1}(d)]. We note that $\Delta_{probe}+\Delta_{wire}(\Phi_0)=$ \SI{335}{\micro\eV}, while $\Delta_{probe}+\Delta_{wire}(\Phi_0/2)=$ \SI{256}{\micro\eV}. Considering for the superconducting gap of the Al tunnel probe a value equal to the textbook value of \SI{180}{\micro\eV}, we can conclude that the fully induced superconducting gap in the wire is \SI{155}{\micro\eV}. The residual gap at $\Phi_0/2$ is \SI{76}{\micro\eV}, which corresponds to a tunability of \SI{79}{\micro\eV}, hence the 51\%. Across both devices tested, we observed similar behavior, with a variation of less than 10\% in all the aforementioned proximity-related features.
Finally, from the IV curves we estimate the tunnel resistances, which, given our fabrication parameters, fall within the range \SI{}{\mega\ohm}. 
In the following discussion, we refer to two specific SQUIPTs, chosen to implement the bi-SQUIPT, with tunnel resistances of \SI{1.18}{\mega\ohm} and \SI{35.71}{\mega\ohm}, respectively.

\section{Bi-SQUIPT voltage swing and sensitivity}
\begin{figure}[t]
    \centering
    \includegraphics[width=1\columnwidth]{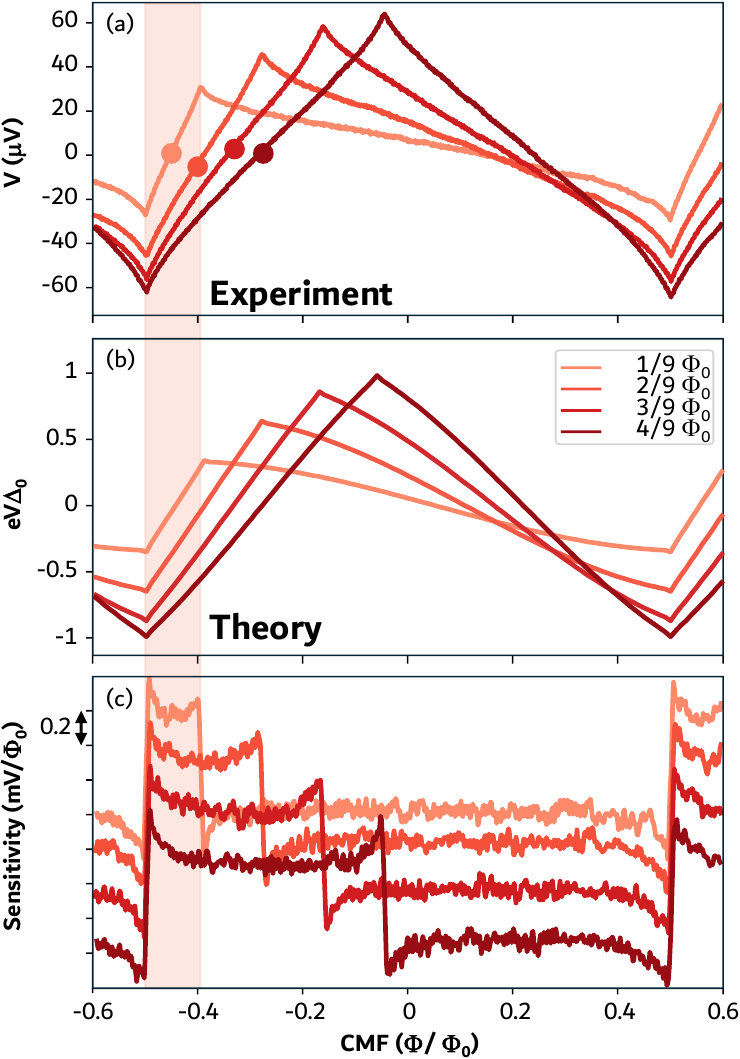}
    \caption{\textbf{Voltage swing and sensitivity of a bi-SQUIPT:} (a) Theoretical normalized voltage response under normalized DC biases  $I_{b,x}=0.1 \Delta_0 / e R_{x}$ of an ideal (same loop areas and tunnel probe resistivity) bi-SQUIPT as a function of the CMF, for selected values of the DMF. These curves were obtained by setting $T=0.01 T_C$, where $T_C$ is the critical temperature of the superconductor, and for a Dynes parameter $\Gamma=10^{-4} \Delta_0$, where $\Delta_0$ is the zero-temperature superconducting energy gap. (b) Voltage swing of a bi-SQUIPT measured at different differential magnetic fluxes as a function of the common mode magnetic field for the same selected values of DMF, as shown in panel a. (c) Sensitivity to magnetic flux of a bi-SQUIPT relative to the curves shown in panel b. The portion of the plot highlighted in red represents the dynamic range for $DMF = 1/9\cdot \Phi_0$. The dots indicate the optimal operating points used to calculate the SFDR.}
    \label{fig:Fig2}
\end{figure}
Since $R_a \not= R_b$, when working in a bi-SQUIPT configuration, two different bias currents must be provided to each SQUIPT. Although theoretically \cite{desimoni2023}, $I_{b,x}$ must be set in the order of $0.1(\Delta_{probe}+\Delta_{wire})/eR_x$ to maximize the voltage swing, we experimentally determined the most convenient bias current for each SQUIPT by measuring the voltage swing of the bi-SQUIPT as a function of both bias currents. 
We found that those that maximize the voltage swing, providing a peak-to-peak modulation amplitude of about \SI{120}{\micro\volt}, \textit{i.e.,} twice that of a single SQUIPT, are $I_{b,a}=$ \SI{55.0}{\pico\ampere} and $I_{b,b}=$ \SI{4.9}{\pico\ampere}. Note that they differ by an order of magnitude, which is consistent with the difference in tunneling resistance of the two SQUIPTs. Note that a bias current always exists that maximizes the voltage swing in a single SQUIPT, regardless of the probe tunnel resistance. This is because the voltage swing depends only on the magnitude of the induced minigap, rather than on the absolute value of $R_x$.  This characteristic makes the bi-SQUIPT very robust against fabrication defects.

We now focus on the flux response of the bi-SQUIPT. For simplicity, we define the differential magnetic flux $DMF=\Phi_{b}-\Phi_{a}$, and the common mode magnetic flux $CMF$. The latter is the magnetic flux induced by the superconducting coil, which is common to both SQUIPTs (the areas of the loops are supposed to be identical), while $DMF$ is independently set for each SQUIPT and controlled by acting on $I_{f,a}$ and $I_{f,b}$.
In \autoref{fig:Fig2}(a), we show the $V-\Phi$ characteristics of bi-SQUIPT as a function of the CMF, for selected values of the DMF. Each DMF was provided by feeding a different DC value in the flux line of one of the SQUIPTs, whereas the CMF was set through the superconducting coil. When the DMF is low, $V-\Phi$ presents a positive-derivative branch of the low-slope dynamic range (highlighted in red in \autoref{fig:Fig2} (a) for $DMF = 1/9\cdot\Phi_0$) and a negative-derivative branch of the low-slope high-dynamic-range. The asymmetry between the positive-derivative and the negative-derivative branch is progressively reduced by increasing the DMF, to the point that, when DMF $\approx0.5\Phi_0$, the two branches are specular with respect to $\Phi=0$. Increasing the DMF further, the behavior of the two branches is inverted.

It is relevant to make a direct comparison between the current–flux experimental characteristic of bi-SQUIPT and the theoretical response expected for an ideal device (with identical loop areas and tunnel resistances), using the approach described in \cite{desimoni2023}. \autoref{fig:Fig2} (b) shows theoretical normalized $V-\Phi$ characteristics calculated for bias currents $I_{b,x}=0.1 \Delta_0 / e R_{x}$ and setting $T=0.01 T_C$, where $T_C$ is the critical temperature of the superconductor, and $\Gamma=10^{-4} \Delta_0$, for selected DMF values. 
It is worth highlighting that the qualitative agreement between the theoretical and experimental curves is excellent. This is due to the robustness of both the model and the bi-SQUIPT against fabrication imperfections, which can be easily compensated by setting two distinct bias currents for the two SQUIPTs.

Given the periodic behavior of the bi-SQUIPT, we can restrict our discussion to the positive-derivative side of the curve, since by acting on the DMF we can readily transition from a high-dynamic-range, low-slope curve to its converse. This suggests that, depending on the specific application, the CMF and DMF working points should be set to prioritize sensitivity or dynamic range, respectively.  This configuration is supported by the data reported in \autoref{fig:Fig2} (c), which shows the sensitivity of the bi-SQUIPT to magnetic flux calculated from the curves shown in panel (a), by evaluating the numerical derivative $\frac{dV}{d\Phi}$ of the data. Here, we note that for lower DMF values, the sensitivity to magnetic fields increases, reaching \SI{0.6}{\milli\volt/}$\Phi_0$, but the dynamic range accessible in this regime is reduced, lasting only about $0.1\Phi_0$. However, as DMF increases, the sensitivity decreases while the dynamic range increases. In a sense, there is a trade-off between dynamic range and sensitivity, and this trade-off should be considered differently across applications.

\section{Transfer function linearity}
\begin{figure}[t]
    \centering
    \includegraphics[width=1\columnwidth]{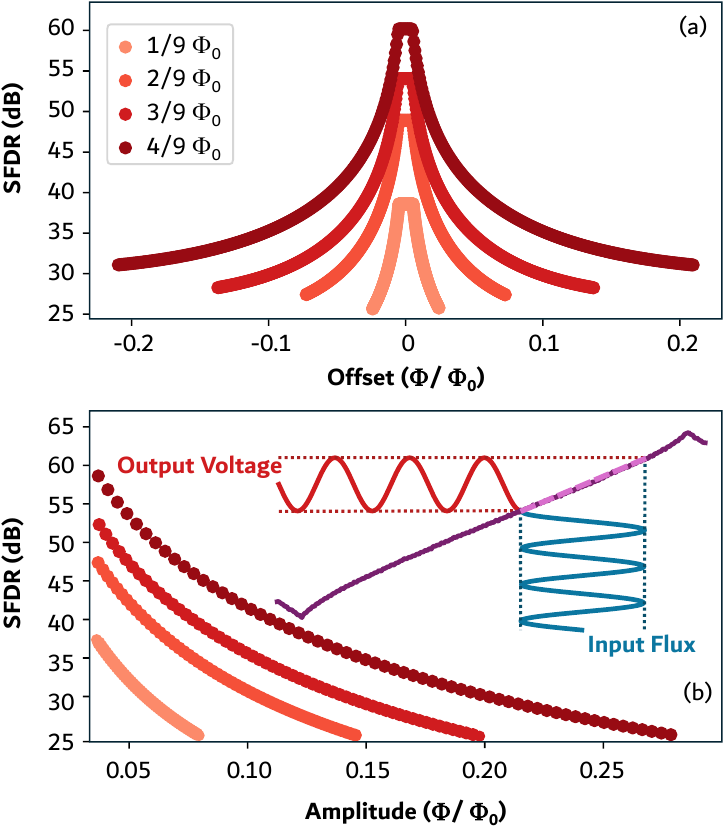}
    \caption{\textbf{Study of the SFDR:} \textbf{(a)} SFDR calculated for fixed input amplitude as a function of the offset from the optimal working points. \textbf{(b)} SFDR calculated at the optimal working points as a function of the input amplitude. Inset: Sketch for the calculation process of the SFDR. A monochromatic wave (blue) is fed into the system, which transforms it via its transfer function (violet), yielding an output signal (red) whose Fourier transform we study. }
    \label{fig:Fig3}
\end{figure}
The Spurious-Free Dynamic Range (SFDR) is a performance metric that quantifies the usable dynamic range of analog systems, particularly in high-frequency applications. It is defined as the ratio, usually expressed in decibels, of the fundamental signal's amplitude to the amplitude of the highest spurious spectral component within the system bandwidth, typically originating from nonlinear distortion or intermodulation. 
SFDR, therefore, characterizes the ability of a transducer or an amplifier to process signals without significant spectral contamination, and a higher SFDR indicates superior linearity and spectral purity.

The SFDR can be evaluated by feeding one or two tones into the system under test and measuring the output spectrum to determine the level of the fundamental and the most prominent generated spurious tone. By calling $A_1$ the amplitude of the basic harmonic and $A_M$ the amplitude of the second-highest harmonic in the Fourier spectrum of the output, we adopt the following definition:
\begin{equation}
    SFDR = -20log\left( \frac{A_M}{A_1} \right)
\end{equation}
The direct measurement of the SFDR of SQUID-based amplifiers is usually performed in the $10-100\:$\SI{}{\mega\hertz} interval. In our case, since the filtered lines for cryogenic characterization have a cut-off frequency of the order of $\approx\:$\SI{}{\kilo\hertz}, we perform an estimation procedure based on the digitalization of the transfer function (TF) (see inset  \autoref{fig:Fig3} (b)), represented by the curves in \autoref{fig:Fig2}(a). We fit the transfer functions $T(\Phi)$ data to spline curves, then calculate the system response to a monochromatic input $I(t)$ by $O(\Phi,t)=T(\Phi)I(t)$ and finally calculate the SFDR on the Fourier transform of $O(\Phi,t)$ (inset of \autoref{fig:Fig3} (a)).

The value of SFDR is strongly dependent on the TF CMF operating point. For this reason, we first calculate the SFDR for a fixed input amplitude ($\approx0.01\Phi_0$) as a function of the working point by sweeping it across the dynamic range of each TF. \autoref{fig:Fig3}(a) shows the calculated SFDR as a function of the offset from the optimal working point of the TF for different DMF. Depending on the DMF, the CMF operating point that maximizes the SFDR varies.  Such sweet spots are indicated for the selected DMF values in  \autoref{fig:Fig2}(a) as thick dots on top of the data points. Although the described technique is not meant to substitute the direct measurement of the system SFDR, we wish to emphasize that the calculated SFDR at the sweet spots can reach a remarkable value of 60 dB. \textit{i. e.,} a value that is exceptionally close to the theoretical values calculated in \cite{desimoni2023}, and highly promising because it is competitive with the linearity performance of SQAs, yet achieved with a double-cell interferometer. When moving from the sweet spot, the SFDR drops dramatically.
In \autoref{fig:Fig3}(b), we show the SFDR calculated for selected DMFs at the sweet spots as a function of the input amplitude. As expected, the linearity of the voltage response strongly depends on the amplitude of the input flux signal, since increasing its amplitude necessarily probes regions near the cusps in the voltage-flux characteristic. Furthermore, it should be noted that, by increasing the DMF in the $0<DMF<\frac{\Phi_0}{2}$ interval, the SFDR also increases, providing a more linear response for a wider range of CMF.

\begin{figure}[t]
    \centering
    \includegraphics[width=1\columnwidth]{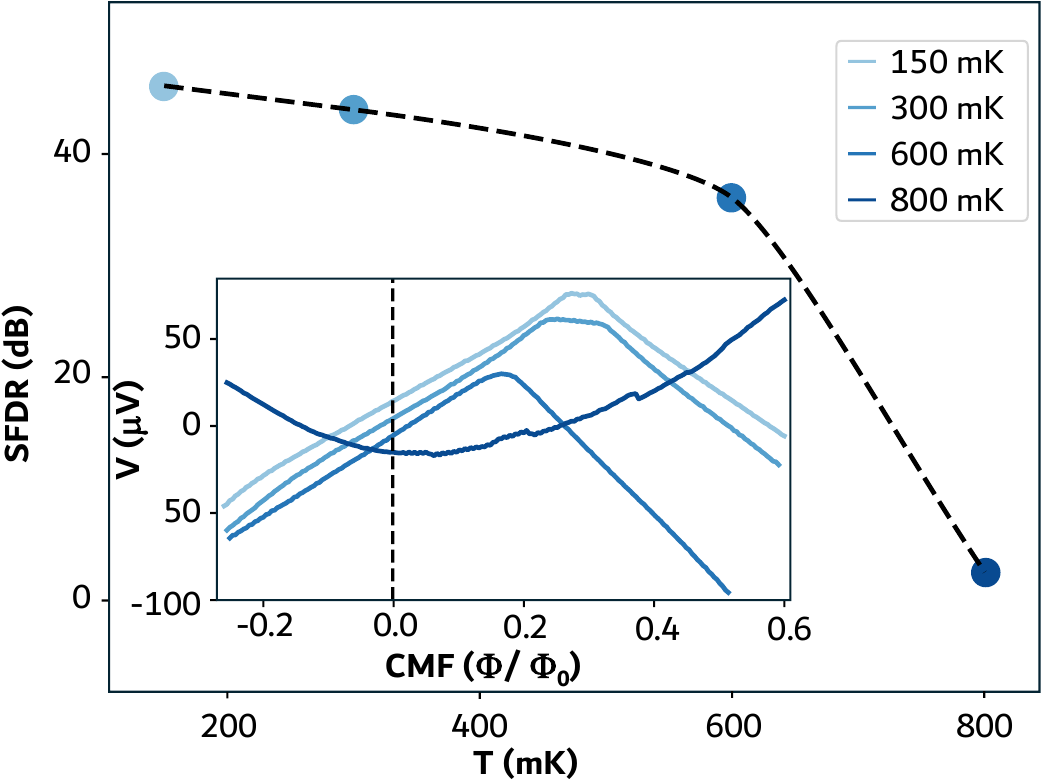}
    \caption{\textbf{Temperature performance:} SFDR calculated by taking the optimal working point at base temperature for $DMF = 4/9\cdot \Phi_0$ as a function of temperature. The dashed black line is a guide for the eye. Inset: TF measured as a function of temperature for fixed DMF. The curves are shifted along the voltage axis for clarity. The vertical dashed line indicates the CMF point where the SFDR is calculated in the inset.
    }
    \label{fig:Fig4}
\end{figure}

To quantify the resilience of bi-SQUIPT to temperature fluctuations, we also acquired the characteristics $V-\Phi$ for selected bath temperatures $T$ (see inset of \autoref{fig:Fig4}). The device response stability is governed by the evolution of the superconducting order parameter in the Al loop and by its consequent ability to efficiently proximitize the weak-link. In fact, $V-\Phi$ remains essentially stable up to 600 mK, which is consistent with a temperature approximately equivalent to one-half the critical temperature of aluminum and above which $\Delta_{Al}$ starts to decay. Above such a temperature, $V-\Phi$ is strongly distorted. This behavior directly impacts the performance of the SFDR, which is computed as a function of $T$ at the sweet spot for $DMF = 4/9\cdot \Phi_0$ and scatter-plotted as a function of $T$ in \autoref{fig:Fig4}. With a best performance value of \SI{46}{\decibel} obtained below 200 mK, the SFDR exhibits a monotonic decrease, which leads to a total loss of linearity at approximately \SI{800}{\milli\kelvin}. Remarkably, the SFDR at \SI{600}{\milli\kelvin} is still \SI{36}{\decibel}.

\section*{Conclusions}
In conclusion, we have realized the first experimental implementation of the bi-SQUIPT, showing that a proximity-effect-based differential architecture can effectively overcome the linearity limitations inherent to superconducting magnetometry. We have shown that the parallel combination of two SQUIPT elements yields a highly linearized transfer function (producing SFDR $\sim 60$ dB below 150 mK) that rivals the performance of much larger superconducting quantum arrays. 
A critical finding of this study is the device's inherent robustness to fabrication variations, as evidenced by our ability to compensate for significant mismatches in tunnel resistances by precisely adjusting individual bias currents.
From a thermal management perspective, the negligible power dissipation and the retention of a spurious-free dynamic range of $\approx 35$ dB at temperatures up to 600 mK make the bi-SQUIPT an ideal candidate for integration into the ultra-low-temperature stages of dilution refrigerators. 
The flexibility afforded by independent control of common-mode and differential flux enables a versatile trade-off between sensitivity and dynamic range, accommodating the specific needs of diverse applications. Ultimately, this work validates the bi-SQUIPT as a compact, ultra-low-power, and high-linearity sensor, paving the way for advanced qubit readout systems and cryogenic diagnostics where both spectral purity and high integration density are paramount.

\section*{Acknowledgments}
The authors thank the PNRR MUR project PE0000023-NQSTI for partial financial support.

\end{document}